\documentclass[usenatbib]{basi}
\usepackage[T1]{fontenc}
\usepackage[british]{babel}
\usepackage[varg]{txfonts}
\usepackage{epsfig}
\usepackage{longtable}
\usepackage{rotating}
\usepackage{dcolumn}

\def\deg{$^\circ$}

\def\aap{A\&A}

\def\apj{ApJ}

\def\aj{AJ}

\def\mnras{MNRAS}
\def\pasp{PASP}

\begin{document}
\title[BASI sample paper]{Near-infrared properties of classical novae: a perspective gained  from Mount Abu Infrared Observatory }
\author[D.P.K.~Banerjee \& N.M.~Ashok]%
       {D.P.K.~Banerjee\thanks{email: \texttt{orion@prl.res.in}},
       and N.M.~Ashok\thanks{email: \texttt{ashok@prl.res.in}}\\
       Physical Research Laboratory, Navrangpura, Ahmedabad, Gujarat 380009, India}
\pubyear{2012}
\volume{00}
\pagerange{\pageref{firstpage}--\pageref{lastpage}}

\date{Received --- ; accepted ---}

\maketitle
\label{firstpage}

\begin{abstract}

 We review the near-infrared properties of classical novae in the $J$, $H$ and $K$ bands
 at wavelengths between 1.08 to 2.4 $\mu$m.  A classification system exists for the early post-outburst optical spectra of novae on the basis of the strength of group of non-hydrogen emission lines. A similar scheme for the near-infrared regime, which is not available at present,  is presented here.  In the optical system
there are two principal classes, namely, "Fe II" and "He/N" for novae with either prominent Fe II lines or
prominent "He/N" lines. There is also a small subset of the  hybrid Fe IIb type. From spectroscopic observations we show the differences and similarities between
these  classes  of novae in the near-infrared. The spectral lines common to the two principal classes arise from H, He, N and O. However, the  near-IR features that separate these two classes are the numerous, and often strong,  Carbon lines which are seen only in the spectra of the  Fe II class of novae. The dust formation process in novae is discussed based on broad-band observations. The  first-overtone carbon monoxide (CO) detections in novae are analyzed to understand the formation and evolution of this molecule in the nova ejecta and to discuss the observed $^{12}$C/$^{13}$C ratio.

\end{abstract}



\section{Introduction:}
\subsection{A brief preamble to the nova phenomenon} The cause and origin of the nova outburst has been described in  detail in other accompanying articles in the present compendium as also in several review articles in Bode $\&$ Evans (2008). We thus do not elaborate further on this but paraphrase the essentials from the article by Chesneau $\&$ Banerjee (2012) in the present volume. A classical nova eruption results from a
thermonuclear runaway (TNR) on the surface of a white dwarf (WD) that is accreting material from a companion star in a close binary system. The accreted hydrogen rich material gradually forms a layer on the WD's surface with the mass of the layer growing with time. As the accreted matter is  compressed and heated by the gravity of the WD,  the critical temperature and pressure  needed for the TNR  to commence are  reached in the degenerate material at the base of the accreted layer thereby starting  the nova eruption.  Observationally, the outburst is accompanied by a large brightening, generally with an amplitude of 7 to 15 magnitudes above the brightness of the object in quiescence. The subsequent development of the nova, in different wavelength regimes, is described in Bode $\&$ Evans (2008); a comprehensive general  overview of  Cataclysmic Variables and the nova phenomenon is given in Warner (1995).

\subsection{Scope of this work}
A classification system for the early post-outburst optical spectra of novae has been devised
by Williams (1992) on the basis of the strength of group of non-hydrogen emission lines. In this system
there are two classes, namely, "Fe II" and "He/N" for novae with either
prominent "Fe II" lines or
prominent "He/N" lines. The Fe II novae have lower level of ionization, display P Cygni absorption components
and their temporal evolution is slow. In contrast, He/N novae have higher level of ionization with the
lines that are flat topped with little absorption and exhibit large expansion velocities. The temporal
evolution of He/N novae is very fast leading to the appearance of coronal lines and strong neon lines.
The classification scheme of Williams (1992) is widely followed in the optical.

{\it However, a similar classification  scheme has not been attempted in the near infrared. We have been observing novae from the Mt. Abu observatory for slightly over a decade and have obtained  spectra of nearly 40 novae during this time. Because of the modest 1.2 meter diameter of the telescope, these spectra are mostly limited to the early period after the outburst when the nova is bright --- a  period generally  extending a few weeks to a few months after the eruption. Based on such early spectra, we show and discuss how the spectra of Fe II and He/N novae appear  in the near-infrared and the similarities and differences that are observed between them. That is, we  extend the optical classification scheme into the  near-IR specrtal region; this is the central and major goal of this work.} The near-spectra that we use are obtained in the $J$,$H$ and $K$ bands covering the 1.08 to 2.35 $\mu$m region. The IR observations of novae beyond 3$\mu$m are reviewed in a companion paper in this issue by Evans $\&$ Gehrz (2012).

We also examine the dust formation process in novae showing its influence on the near-IR light curves of selected novae. In the context of dust formation, we also propose a simple predictive scheme to foretell which novae will form dust which can easily be proved true or false. Finally we discuss CO formation and evolution in novae based on the $K$ band detection of CO emission in the first overtone. Only a few such $K$ band detections are recorded in novae and two of these were detected and studied in considerable detail from Mt. Abu.

We do not discuss the near-IR properties of recurrent novae here though the most recent outbursts of four of these viz.,  CI Aql, RS Oph, U Sco and T Pyx have been studied from Mt. Abu (the last three RNe were infact studied extensively ;  Das et al. 2006, Banerjee et al. 2009, Banerjee et al. 2010, Banerjee $\&$ Ashok 2011, Chesneau et al. 2011).  We also do not discuss interesting "nova-like" objects like V838 Mon (Banerjee $\&$ Ashok 2002, Banerjee et al. 2005) or  V4332 Sgr which have been pursued by us or the nova V445 Puppis which was proposed to be the first helium nova based on Mt Abu observations (Ashok $\&$ Banerjee, 2003).

\subsection{Mount Abu Observatory}
Mount Abu is the highest hill station, with Gurusikhar its highest peak at an altitude of 1722 m, of the Aravalli ranges of central India  that lie in the middle of  the  $\sim$ 2000 km N-S extent of India extending  from the Himalayas in the North  and the blueness-endowing,  eucalyptus-clad Nilgiris (Blue Mountains) in the south. Mt Abu observatory is geographically situated at 78\deg E, 24\deg N,  221 kms from Ahmedabad,   the scientific headquarter of the Physical Research Laboratory that runs the observatory.  The distance between the two places  is covered in a 4 to 5 hour drive by road, mostly over dry, hot plains except for the last 20 kms when  the ascent up the hills begin. Notable landmarks on the way are few but one may count  the spiritual shrine of Ambaji with its glittering gilt-domed temple while the journey is spiced up  to some c extent by  the fennel-cumin-isabgol growing "spice capital" of Unjha on the way. A  thrill is  also invariably  felt in crossing the Tropic of Cancer that must inevitably be crossed on the way to reach Abu. The observatory, with its  pleasant weather is an extremely enticing retreat from Ahmedabad which informally has only two seasons - a summer (with the temperature in the range 35-40\deg C) and a very hot summer (40-45\deg C plus) -  a place where it is likely to see sweat roll down the knose, kneck and knees, down the shins and into one's shoes  to emerge as steam from the  lace holes (apropos Spike Milligan).

Except for the rainy  months of  July, August and September  which is the monsoon period during which most tracts of India receive their complete annual quota of rainfall,   the observatory is open for the rest of the year (the annual rainfall at Mt. Abu varies  between 100 to 150 cms).  Inspite of the  hiatus in observations during the monsoon,  Mt. Abu boasts of  an enviable number of clear nights and it is rare for an observer to be totally clouded out.  A systematic survey of the INSAT satellite cloud imagery database over the period 1989-94 shows that the annual percentage of clear (spectroscopic) nights  is 70$\%$ - the national best among the Indian sites surveyed (Sapru et al. 1998) . The site also has good seeing which has been typically measured in the optical to be sub-arcsecond from DIMM  (Ashok $\&$ Banerjee, 2002) and some SIMM (Differential and Single Image motion Monitors respectively) measurements as well as direct FWHM measurements of star images. The seeing in the near-IR is thus expected to be even better since the Fried parameter increases with wavelength resulting in the seeing  FWHM varying as  $\lambda^{-1/5}$.

Standing alone on the observatory summit, looking down the dizzying basalt precipices, with the serene tolling of the bells of the Saint Dattatreya temple situated on the neighboring peak filling the air with a caress of holiness, the human spirit  is metaphorically close to finding its own  Walden here. The soul is lifted to a high degree of introspection and one almost  "ceases to live and begins to be" (Henry David Thoreau, "Walden"). As Thomas Hardy,  a keen observer of the skies  who could readily distinguish  "the steely glitter and sovereign brilliancy of a Sirius from the fiery red of Aldebaran and Betelgueux",  wrote of a dark sky on a dark night, it is sometimes " necessary to stand on a hill at a small hour of the night, and, having first expanded with a sense of difference from the mass of civilised mankind, who are dreamwrapt and disregardful of all such proceedings at this time, long and quietly watch your stately progress through the stars. After such a nocturnal reconnoitre it is hard to get back to earth, and to believe that the consciousness of such majestic speeding is derived from a tiny human frame." (Thomas Hardy, "Far from the madding crowd")

The return after a night's observations,  especially in the wee hours when it is still dark, allows one to see plenty  of wildlife. The occasional bear can be seen as too an adult porcupine  bristling aggressively with all quills stiffly erect. Wild hares abound,  evincing empathy as they race desperately along the road following the footprints of the headlights,  just yards ahead of the wheels,  till a bend in the road or sheer exhaustion forces  them to veer off the road and out of harm's way.  However, the local Rajasthani drivers are careful - including the observatory jeep driver - and rarely does one see a road kill. But the experience that is unsurpassable in impact is the sighting  of the  leopard - an event that occurs quite frequently. Sometimes good fortune allows even a male and female pair of them to be seen together,  crossing the road ahead, unafraid of the approaching vehicle; mocking our anthropic sense of superiority through their languid, unhurried and majestic walk across the road.

\section{Instrumentation}
The astronomy division has since inception  developed  a suite of instruments for observational use. These include Fabry Perot spectrometers, grating spectrographs, polarimeters  and an echelle spectrograph. However, most of the nova studies described here have been done with a custom-made near-IR imager spectrograph. that has been in use since 1998.  The instrument, which  employs a 256x256 HgCdTe NICMOS-3 detector with 40 micron pixels, can record images or spectra in the near-IR $JHK$ bands. The optics is essentially  cadoptric consisting of the collimating and camera mirrors arranged in a Ebert Fastie configuration and a 149 lines/mm grating. The only refractive element is a focal-reducer lens, whose use is optional,  placed before the slit. The rearside of the grating is silvered to act as a mirror and the grating-cum-mirror assembly is mounted on a rotatable platform to allow  switching between imaging and spectroscopy modes.  The instrument is designed to give a similar dispersion of about 9.5\AA ~per pixel in each of the $J$,$H$ and $K$ bands, which when coupled to the projected width of the 76 $\mu$m slit to 2 pixels on the detector,  gives an average spectral resolution of $\sim$ 1000 in the $JHK$ bands.

\section{How  Fe II and He/N novae differ in their near-infrared spectra}

\subsection{Principal spectral characteristics and line identification}

The representative  $JHK$ spectra for four  Fe II novae and four He/N novae are shown in Figures 1, 2 and 3.  These spectra represent the phase when the P Cygni phase is over and  the nova lines are clearly in emission. The earliest spectra, when the P Cygni absorption components are prominent are not shown here but these can be seen for the novae in Figures 1, 2 and 3 in the published references associated with them. The Fe II novae show a larger number of lines vis-s-vis the He/N novae and most of these lines, using V1280 Sco as a typical example,  are identified in Figure 4 and also listed in Table 1. The lines generally seen in the early spectra of He/N class of novae are listed in Table 2.

The principal differentiating feature between the two classes is the strong presence of Carbon lines in Fe II novae and their absence in He/N novae. This is a key result. Otherwise both classes are similar in showing strong Hydrogen lines of the Paschen and Brackett series. Regarding oxygen lines, both classes behave similarly and show the   Lyman-beta fluoresced OI 1.1287 $\mu$m line prominently. The OI 1.3164 $\mu$m line is also seen in both classes -- this line is believed to be excited by a combination of continuum fluorescence and collisional excitation but not by accidental resonance with the Lyman beta wavelength (Bhatia $\&$ Kastner 1995, Kastner $\&$ Bhatia 1995, Blesson et al. 2012). Regarding helium lines, He/N
novae show much stronger He emission right from the start of the outburst. The principal He lines that we detect are the 1.0830, 1.7002, 2.0581 and the 2.1120,2.1132 $\mu$m lines. Few near-IR lines of nitrogen are detected in both Fe II or He/N spectra (for e.g the 1.2461, 1.2469 $\mu$m lines) but they are invariably stronger in He/N novae. In this aspect, the strength of the He and N lines behave similarly with their counterpart optical spectra (Williams 1992).

{\it Thus the near-IR features that instantly allow to distinguish between the spectra of the two classes of novae are the Carbon lines.} The strongest of these C lines are in the $J$ band at 1.166 and 1.175 $\mu$m and in the $H$ band at 1.6890 $\mu$m and several lines between 1.72 to 1.79 $\mu$m. Note that the $K$ band also has several C lines but these are weaker.

If one wishes to see a pure carbon spectrum in a nova i.e. how the carbon lines in a nova spectrum would look like without being adulterated or confused by neighboring hydrogen or oxygen lines, then this can be seen in V445 Puppis. The spectrum of V445 Pup was devoid of H lines and  rich in C and He lines (Ashok $\&$ Banerjee 2003 and references therein).

     The C-rich spectra of Fe II novae may have its origin in a difference of hardness of the radiation field of the central remnant in the two classes of novae just after the nova explosion. He/N novae may have slightly hotter central remnants compared to Fe II novae with a correspondingly larger flux of UV photons below the carbon ionizing continuum at 1102 $\AA$ (equivalently 11.26 eV). Through photo-ionization, the radiation field  would then effectively  decrease  the neutral carbon content responsible for the CI lines in the spectrum. This hypothesis, that ionization and excitation effects are at the origin of the CI lines, needs to be confirmed through rigorous modeling using  appropriate photo-ionization codes; a process which we are pursuing. Pontefract $\&$ Rawlings (2004) have shown how molecule and dust formation in novae needs the presence of a neutral carbon curtain to shield the molecule/dust forming cooler regions from the incident photo-ionizing flux ( a discussion on this is given at the beginning of section 5.1).  The strong tendency of dust to preferentially form in Fe II novae, and rarely in He/N novae,   is consistent with the existence of a neutral carbon curtain in Fe II novae. The differing behavior of the CI lines in the two classes of novae could possibly be explained in this manner.

\begin{figure}[!htp]
\center
\includegraphics[bb=3 18 411 527, width=5in,height=6in, clip]{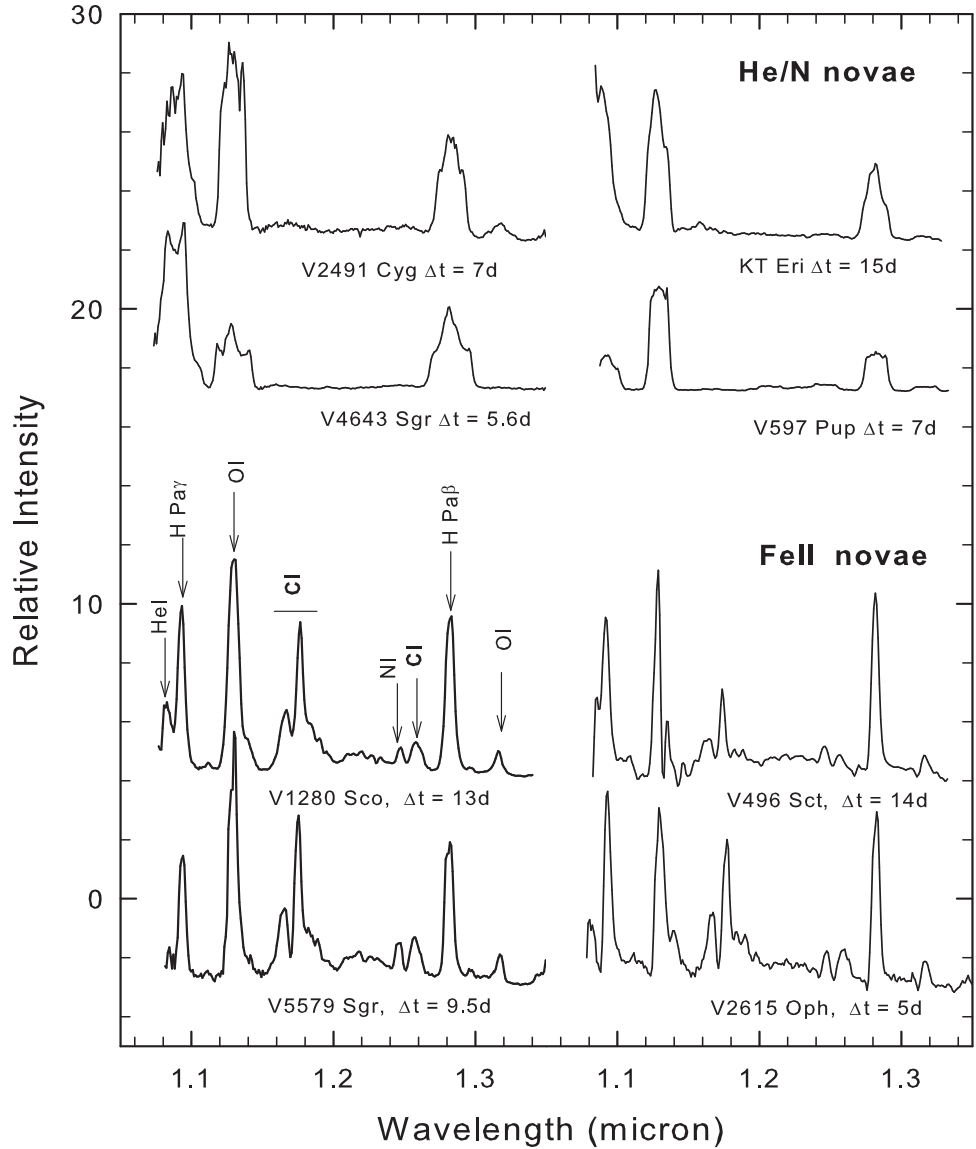}
\caption{The early $J$ band spectra of four novae of the Fe II and the He/N class with the time $\Delta$t after outburst indicated. The major difference between the two classes are the strong Carbon lines seen in the Fe II novae. The other lines are common. The spectra are adapted from: V2491 Cyg and V597 Pup (Naik et al. 2009), V4643 Sgr (Ashok et al. 2006), KT Eri (Raj, Ashok, Banerjee in preparation), V1280 Sco and V2615 Oph (Das et al. 2008 and 2009), V5579 Sgr and V496 Sct (Raj et al. 2011, 2012)}
\end{figure}

\begin{figure}[!htp]
\center
\includegraphics[bb=3 18 419 530, width=5in,height=6in, clip]{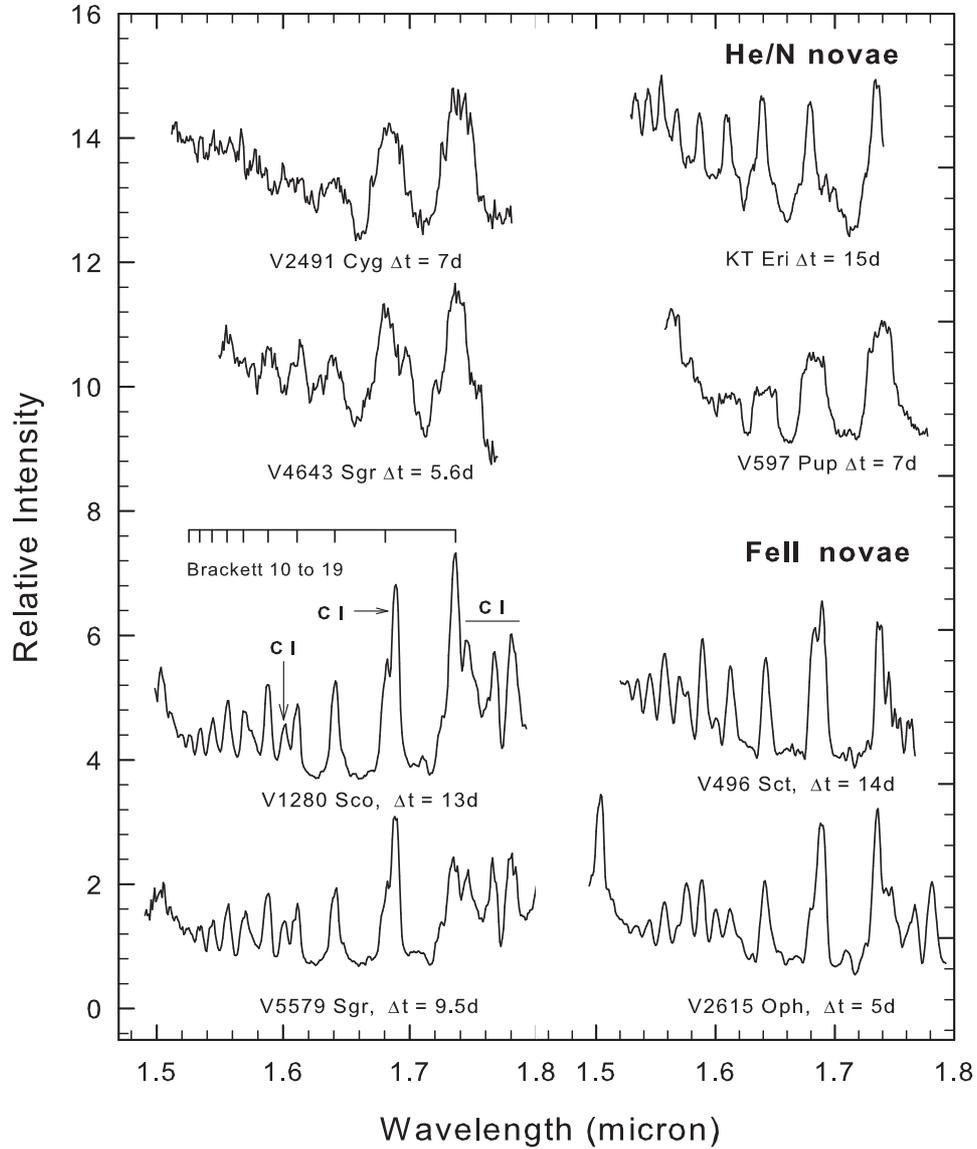}
\caption{The early $H$ band spectra of four novae of the Fe II and the He/N class with the time $\Delta$t after outburst indicated. The lines common to both classes are the hydrogen Brackett series lines. The major difference are the prominent Carbon lines seen in the Fe II novae in the 1.69 to 1.8 $\mu$m region. The spectra are adapted from: V2491 Cyg and V597 Pup (Naik et al. 2009), V4643 Sgr (Ashok et al. 2006), KT Eri (Raj, Ashok, Banerjee in preparation), V1280 Sco and V2615 Oph (Das et al. 2008 and 2009), V5579 Sgr and V496 Sct (Raj et al. 2011, 2012)}
\end{figure}

\begin{figure}[!htp]
\center
\includegraphics[bb=3 18 418 530, width=5in,height=6in, clip]{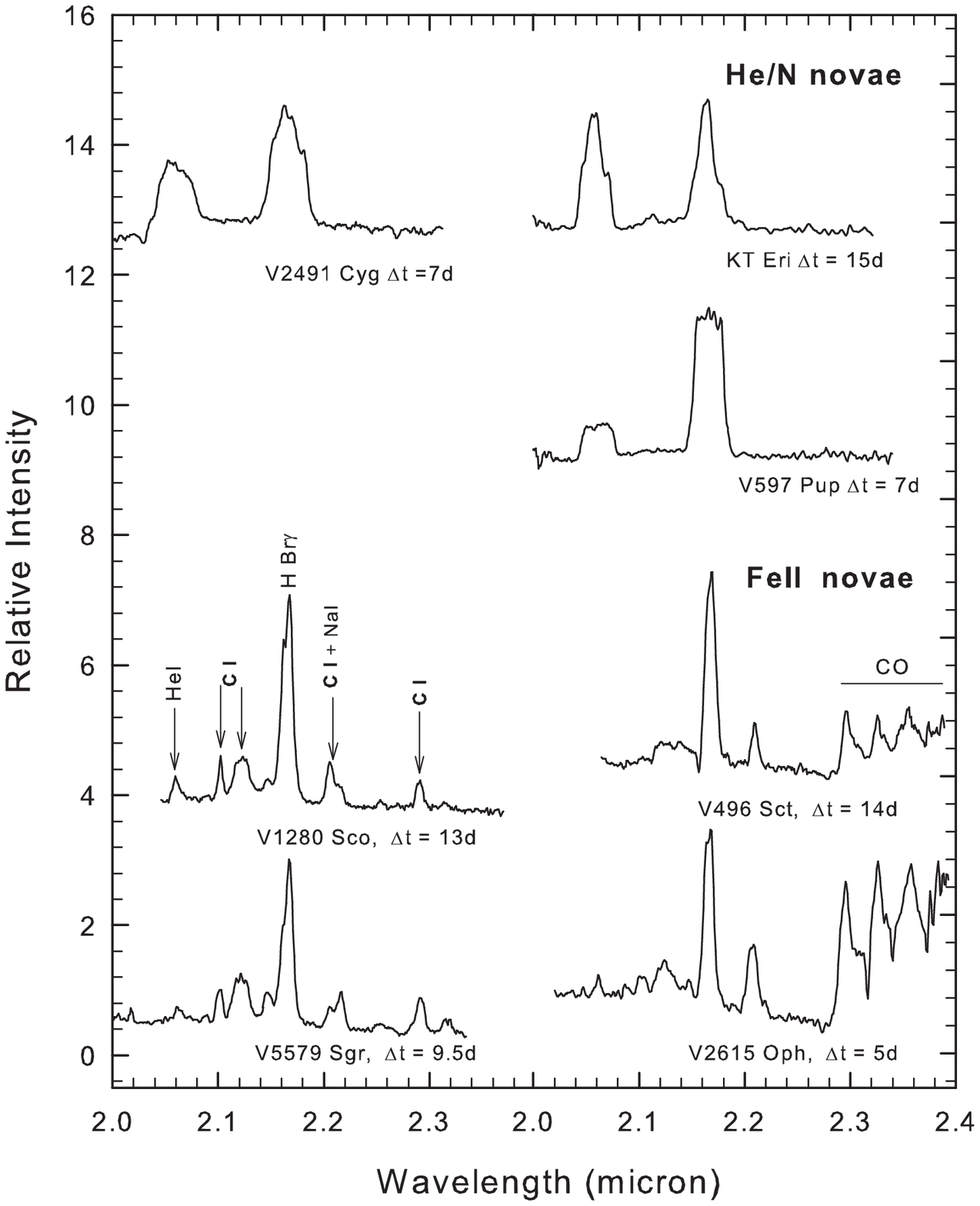}
\caption{The early $K$ band spectra of novae of the Fe II and the He/N class with the time $\Delta$t after outburst indicated. The only strong lines in the He/N novae are the HeI 2.0581, 2.1120, 2.1132 $\mu$m and Brackett $\gamma$ at 2.1655 $\mu$m. The lines in the Fe II novae are identified in more detail in Table 1.  The spectra are adapted from: V2491 Cyg and V597 Pup (Naik et al. 2009),  KT Eri (Raj, Ashok, Banerjee in preparation), V1280 Sco and V2615 Oph (Das et al. 2008 and 2009), V5579 Sgr and V496 Sct (Raj et al. 2011, 2012)}
\end{figure}

\subsection{Finer features of the spectra}

There are several finer features in the spectra that are worth mentioning. Some of these are related to NaI and MgI features which we believe are precursors of dust formation (Das et al. 2008). It should also be pointed out that  the spectra of novae when extended to 0.8 microns show additional emission lines  of HI, HeI, CI, NI, OI and [NI] (Ederoclite et al 2006, Lynch et al 2006, Schwarz et al 2007) and also certain unidentified IR lines (Rudy et al. 2003 and references therein). A few points regarding the spectra presented here:

1) In the $J$ band there is a Mg I line at 1.1828 ${\rm{\mu}}$m in the wing of the strong Carbon lines at $\sim$ 1.1750 ${\rm{\mu}}$m. This Mg I line, in case the nova emission lines are broad in general, may blend  with the C I feature at 1.1750 ${\rm{\mu}}$m giving the latter a broad redward wing. Alternatively it may be  seen as a distinct spectral feature as  seen in V1419 Aql (Lynch et al. 1995) or in  our observations on V1280 Sco  (e.g. the spectra of 3 and 5 March, Das et al 2008). Establishing the identity of Mg lines unambiguously, and also that of Na lines, is important because we show subsequently that they are potential predictors of dust formation.

2) The region between 1.2 to 1.275 ${\rm{\mu}}$m is a complex blend of a very large number of  lines - principally those of NI and C I. This complex  is routinely seen - always  at  low strength - in the early spectra of the Fe II novae that we have studied and those by others  for e.g. V2274 Cyg and V1419 Aql (Rudy et al. 2003; Lynch et al. 1995)

3) The blended feature comprising of the O I 1.1287 ${\rm{\mu}}$m and C I 1.1330 ${\rm{\mu}}$m lines can show considerable evolution with time  as the O I
line component begins to dominate  possibly due to a increase of the Ly $\beta$ flux with time as the central remnant becomes hotter.  The 1.1287 ${\rm{\mu}}$m O I - C I 1.1330 ${\rm{\mu}}$m blended feature generally shows a broad red wing at 1.14 ${\rm{\mu}}$m. There appears to be strong grounds  to believe that Na I lines at 1.1381 and 1.1404 ${\rm{\mu}}$m are responsible for this red wing.  It is also noted that these same Na I features are seen in the spectrum of   V2274 Cen and identified by Rudy  et al. (2003) as potentially arising from Na I.

4) In the $H$ band, the recombination lines of the Brackett series are most prominent and
readily identifiable. But the presence of a C I line at 1.6005 ${\rm{\mu}}$m, which
could  be mistaken as just another member of the Brackett series, should not
be missed. If a recombination analysis of the Hydrogen lines is to be done, caution
is needed in estimating line strengths of Br 11 which can be severely blended with
the strong C I feature at 1.6890 ${\rm{\mu}}$m and Br10 at  1.7362 ${\rm{\mu}}$m which is again blended
with C I lines. Other Br lines, whose strengths are affected but to much lesser extent,
are Br14 at  1.5881 ${\rm{\mu}}$m (blended with C I 1.5853 ${\rm{\mu}}$m and O I 1.5888 ${\rm{\mu}}$m ) thereby making it appear
artificially stronger than Br12 contrary to what is expected in a Case B scenario;
Br12 at 1.6407 ${\rm{\mu}}$m is contaminated with several weak C I lines between
1.6335 ${\rm{\mu}}$m and 1.6505 ${\rm{\mu}}$m thereby
giving Br12 a broad appearance on both  wings. A specifically interesting feature
at 1.579 ${\rm{\mu}}$m  on the redward flank of Br15 (1.5701 ${\rm{\mu}}$m  is due to Mg I;
magnesium and sodium lines are discussed subsequently.\\

5) Apart from the 1.1828 ${\rm{\mu}}$m  line, the other Mg I lines that we  detect are
those at 1.5040,
1.5749 and 1.7109 ${\rm{\mu}}$m . The 1.5040 ${\rm{\mu}}$m and 1.7109 ${\rm{\mu}}$m identifications are also suggested by Rudy et al (2003)
to be due to Mg I.  The
 feature at 1.5749 ${\rm{\mu}}$m, blending with
the wings of Br 15 is certainly Mg I and may escape detection at lower resolutions where it
could blend with lines of the Brackett series. The 1.5749 ${\rm{\mu}}$m line can become quite strong, stronger than the adjacent Br lines as in the case of V2615 Oph (Das et al. 2008) and a
correlated increase in the strength between this line and other Mg I lines is also
clearly seen in V2615 Oph. Regarding lines
arising from Na, apart from those at $\sim$ 1.4 ${\rm{\mu}}$m , the  other lines that are
definitively identified are those at 2.2056 and 2.2084 ${\rm{\mu}}$m . These lines are known
to occur in novae e.g in V705 Cas (Evans et al. 1996). However identifying the 2.1045 ${\rm{\mu}}$m feature
with  Na I is somewhat uncertain.

6) The  C I feature at $\sim$ 2.12 ${\rm{\mu}}$m  is actually a blend of several C I
lines; the principal ones being  at 2.1156, 2.1191, 2.1211, 2.1260 and
2.1295 ${\rm{\mu}}$m.The superposition of these closely spaced lines gives the
overall feature its unusually broad appearance.

\begin{figure}[!htp]
\center
\includegraphics[bb=30 16 270 590, width=3.5in,height=6.15in, clip]{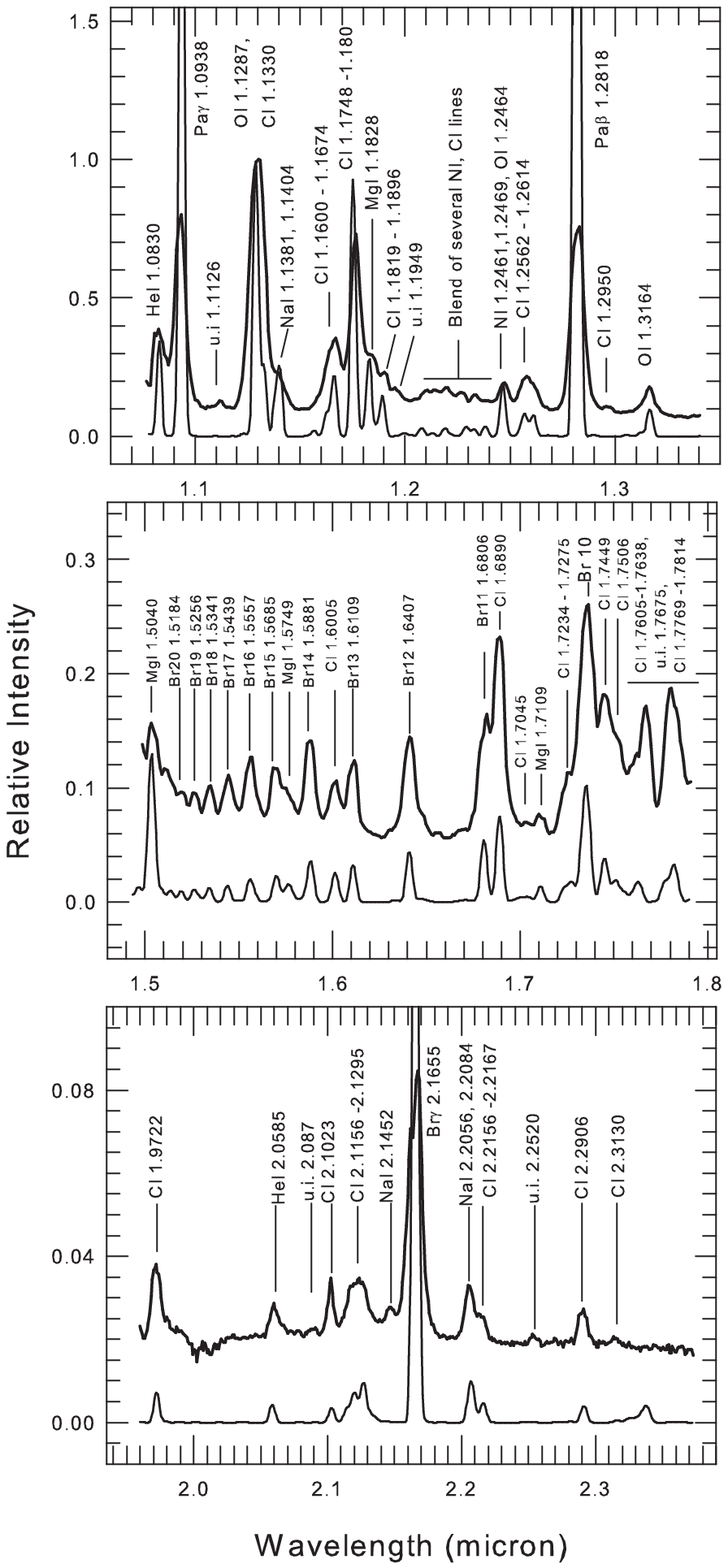}
\caption{Identification of the lines typically seen in a Fe II nova using V1280 Sco as an example.
 The $J,H$ $\&$ $K$ bands shown from top to bottom
respectively. In each panel, the upper plot (darker shade) is the observed data; the lower curve (lighter shade)  is a synthetic spectrum (Das et al. 2008) generated under LTE conditions showing the expected positions and strengths of different lines to facilitate identification. The absorption features at $\sim$ 2.0 ${\rm{\mu}}$m are residuals from improper telluric
subtraction}
\end{figure}

\begin{center}
\begin{longtable}{|l|l|l|l|}
\caption[List of lines commonly seen in the early $JHK$ spectra of Fe II novae]{List of lines commonly seen in the early $JHK$ spectra of Fe II novae} \label{nolabel} \\

\hline \multicolumn{1}{|c|}{Wavelength} & \multicolumn{1}{c|}{Species} & \multicolumn{1}{c|}{Energy (eV) of} & \multicolumn{1}{c|}{Other contributing} \\
 \multicolumn{1}{|c|}{(${\rm{\mu}}$m)} & \multicolumn{1}{c|}{} & \multicolumn{1}{c|}{Lower - Upper levels} & \multicolumn{1}{c|}{ species \& remarks}\\ \hline

\endfirsthead

\multicolumn{4}{c}%
{{\tablename\ \thetable{} -- continued from previous page}} \\
\hline \multicolumn{1}{|c|}{Wavelength} &
\multicolumn{1}{c|}{Species} &
\multicolumn{1}{c|}{Energy (eV) of} &
\multicolumn{1}{c|}{Other contributing} \\
\multicolumn{1}{|c|}{($\mu$m)} &
\multicolumn{1}{c|}{} &
\multicolumn{1}{c|}{Lower - Upper levels} &
\multicolumn{1}{c|}{species \& remarks}\\ \hline
\endhead

\hline \multicolumn{4}{|l|}{{* Range of values for the strongest lines.~~~~~~~~~~~~~~~~~~~~~ Table continued on next page}} \\ \hline
\endfoot

\hline \hline
\endlastfoot

1.0830             	& He \,{\sc i}        	& 19.821 - 20.965 &				\\	
1.0938   	   	& Pa $\gamma$     	&12.088 - 13.222& 		 \\
1.1126   		& u.i             &  & unidentified, Fe \,{\sc ii}? 	\\
1.1287   		& O \,{\sc i}      	& 	10.989 -   12.088&			\\
1.1330   		& C \,{\sc i}      	& 8.538 -  9.632 &\\
1.1381   		& Na \,{\sc i}     	& 2.102 -  3.192& C I 1.1373  \\
1.1404   		& Na \,{\sc i}     	& 2.105 -  3.192 &C I 1.1415  \\
1.1600-1.1674 		& C \,{\sc i}   &  8.640 - 9.830*  	& strongest lines at \\
                &   &  &     	1.1653, 1.1659,1.16696		 \\
1.1748-1.1800 		& C \,{\sc i}& 8.640 - 9.700* 	& strongest lines at 	\\
                        &   &              	&1.1748, 1.1753, 1.1755		\\
1.1828			& Mg \,{\sc i}           & 4.346 -   5.394&                             \\
1.1819-1.1896  		& C \,{\sc i} & 8.640 -  9.690* &strongest lines at             \\
			&			   &  &1.1880, 1.1896          	\\
1.1949              & u.i   &        &  unidentified           \\
1.2074,1.2095   	& N \,{\sc i} &12.01 - 13.04, 12.97 - 13.99   &blended with C I 1.2088     \\
1.2140   		& u.i &           &  	unidentified					\\
1.2187 		& N \,{\sc i}    &11.845 - 12.862 & blended with NI 1.2204                        \\
1.2249,1.2264  		& C \,{\sc i}    &9.70 - 10.72, 9.71 - 10.72 	&					 \\
1.2329   		& N \, {\sc i}  & 11.996 -  13.002&  \\
1.2382   		& N \,{\sc i}  & 11.996 -  12.998 &  \\
1.2461,1.2469  		& N \,{\sc i}& 12.00 - 12.99, 12.01 - 13.00 &blended with O I 1.2464 \\
1.2562,1.2569  		& C \, {\sc i}& 8.849 - 9.835  &blended with O I 1.2570    \\
1.2601,1.2614   	& C \, {\sc i} & 8.851 - 9.835 &    \\
1.2659   		& u.i &           & unidentified \\
1.2818   		& Pa $\beta$ &12.088 -  13.055& 				\\
1.2950   		& C \,{\sc i}& 9.632 - 10.589 &				\\
1.3164   		& O \,{\sc i}		&10.990 - 11.931& \\                         1.5040          & Mg \,{\sc i}&5.108 - 5.932 & blended with Mg I \\
                    & &                      &1.5025,1.5048    \\
1.5192  		& Br 20 &12.749 - 13.565    & \\   		
1.5261  		& Br 19 &12.749 - 13.562&    				\\
1.5342   		& Br 18 &12.749 - 13.557 &	\\
1.5439   	   	& Br 17 &12.749 - 13.552	&  				\\
1.5557   		& Br 16 &12.749 - 13.546 	&   				\\
1.5701  		& Br 15    &12.749 - 13.539&   \\
1.5749              & Mg \,{\sc i} &5.932 -   6.719 &         blended with  Mg I \\
                    &  &                     &1.5741,1.5766, C I 1.5784 \\
1.5881   		& Br 14   & 12.749 -      13.530&           	 blended with C I 1.5853				 \\
1.6005  		& C \,{\sc i}           & 9.632 - 10.406&				\\
1.6109    		& Br 13    		&  12.749 - 13.519& 				\\
1.6335   		& C \,{\sc i}& 9.762 - 10.521 	&  				\\
1.6419  		& C \,{\sc i}& 9.331 - 10.086	&  				\\
1.6407  		& Br 12&  12.749 - 13.505 		&blended with  C I lines \\
                    &  &                &between 1.6335-1.6505\\
1.6806   		& Br 11&12.749 -  13.487   		&  				\\
1.6890   		& C \,{\sc i}& 9.003 -   9.737  	&  				\\
1.7045   		& C \,{\sc i} &9.835 - 10.563  	&  				\\
1.7109                  & Mg \,{\sc i}&5.394 -  6.119   	&  				\\
1.7200-1.7900  		& C \,{\sc i}& 9.690 - 10.410*      	&Several C I lines in 		 \\
			& &			&this region (see Fig.4) 			\\
1.7362  		& Br 10 &12.749 - 13.463 		&Affected by C I 1.7339 		\\
			&		&	&line 				\\
1.9722   		& C \,{\sc i}&  9.003 - 9.632 	&\\
2.0581			& He \,{\sc i}&20.617 -  21.219     &  				\\
2.0870			& u.i  &                 & unidentified 				\\
2.1023 			& C \,{\sc i}&9.172 - 9.762 		& 				\\
2.1138  		& O \,{\sc i} &12.756 -  13.341 	&This line may be present		\\
2.1156-2.1295  		& C \,{\sc i}&9.830 - 10.420*     	&blend of several C I 		\\
                        & &                      &lines strongest being 		 \\
			& &		& 2.1156,2.1191,2.1211, 	\\
			&	&		& 2.1260,2.1295  		\\
2.1452   		& Na \,{\sc i}? &4.284 - 4.861 	&  				\\
2.1655   		& Br $\gamma$ &12.749 -  13.322  	&  				\\
2.2056   		& Na \,{\sc i} & 3.191 -  3.754         & 				\\
2.2084  		& Na \,{\sc i}  &  3.191 -  3.753      & 				\\
2.2156-2.2167	& C \,{\sc i} & 8.770 -9.331*       		&blend of C I lines at 		 \\
			& &			&2.2156,2.2160,2.2167  		\\
2.2520   		& u.i & 			&	unidentified			\\
2.2906   		& C \,{\sc i}& 9.172 - 9.714  	& 				\\
2.3130   		& C \,{\sc i} &10.353 - 10.889	& 				\\
\hline

\end{longtable}
\end{center}

\begin{table}[!htp]
\center
\caption[]{List of lines commonly seen in the early $JHK$ spectra of He/N novae}
\begin{tabular}{llcrc}
\hline\\
Wavelength         	& Species    & Energy (eV) of        	&  \\
(${\rm{\mu}}$m)    	&             & Lower - Upper levels        	&  \\
\hline\\
1.0830          & He \,{\sc i}        	& 19.821 - 20.965 &				\\	
1.0938   	   	& Pa $\gamma$     	&12.088 - 13.222& 		 \\
1.1287   		& O \,{\sc i}      	& 	10.989 -   12.088&			\\
1.1625          & N \,{\sc i}& 10.930 -   11.996 &  \\
1.1651          & N \,{\sc i}& 10.932 -   11.996  &  \\
1.2461  		& N \,{\sc i}& 12.001 - 12.996 &  \\
1.2469  		& N \,{\sc i}& 12.010 - 13.005 &  \\
1.2818   		& Pa $\beta$ &12.088 -  13.055& 				\\
1.3164   		& O \,{\sc i}		&10.990 - 11.931& \\                                 1.3602	        & N \,{\sc i}& 12.123 - 13.034  &  \\
1.3624	        & N \,{\sc i}& 12.127 - 13.037  &  \\
1.5192  		& Br 20 &12.749 - 13.565    & \\   		
1.5261  		& Br 19 &12.749 - 13.562&    				\\
1.5342   		& Br 18 &12.749 - 13.557 &	\\
1.5439   	   	& Br 17 &12.749 - 13.552	&  				\\
1.5557   		& Br 16 &12.749 - 13.546 	&   				\\
1.5701  		& Br 15    &12.749 - 13.539&   \\
1.5881   		& Br 14   & 12.749 -      13.530&           	 \\
1.6109    		& Br 13    		&  12.749 - 13.519& 				\\
1.6407  		& Br 12&  12.749 - 13.505 		& \\
1.6806   		& Br 11&12.749 -  13.487   		&  				\\
1.7002          & He \,{\sc i}        	& 23.009 -  23.738 &				\\	
1.7362  		& Br 10 &12.749 - 13.463 		& 		\\
1.9446	        & Br 8 &12.749 -  13.387   		& 		\\
2.1120          & He \,{\sc i}        	& 23.009 - 23.596&				\\	
2.1132          & He \,{\sc i}        	& 23.089 -  23.675&				\\	
2.0581 			& He \,{\sc i}&20.617 -  21.219     &  				\\
2.1655   		& Br $\gamma$ &12.749 -  13.322  	&  				\\
\hline
\hline \\
\end{tabular}
\end{table}

\subsection{ The near-IR spectra of the  Fe IIb or hybrid class of novae}
In the study by Williams (1992) a small fraction of novae display characteristics of both classes and these are
referred to as "hybrid" or "Fe IIb" novae. These novae have strong Fe II lines soon after the
outburst which are broader than those seen in typical "Fe II" novae and subsequently display "He/N"
spectrum. The "He/N" novae are not seen to evolve to "Fe II" novae. From our sample, we identify V574 Pup to be a representative member of this hybrid class.

 V574 Pup showed a fairly rapid change of its IR spectra from Fe II to He/N class. This is illustrated in Figure 5 wherein the CI lines, characteristic of the Fe II class,  quickly disappear and a remarkable and rapid increase  in the strength of the HeI lines (e.g. the 1.0830 and 2.0581 $\mu$m lines) is seen as expected in the He/N class. A larger mosaic of spectra showing the evolution in detail is given in Naik et al (2010). V574 Pup was discovered to be in outburst
on 2004 November 20.67 UT and low dispersion optical spectra of the nova
on 2004 November 21.75 UT showed H$\alpha$ and H$\beta$ emission
lines with P Cygni components, along with the strong Fe II
(multiplet 42) in absorption indicating that V574 Pup is a
Fe II class nova near maximum light (Ayani 2004).  The
optical spectra of the nova on 2004 November 26 and December 12 obtained by Siviero et al. (2005) were found to
be dominated by Balmer hydrogen and Fe II emission lines;
no nebular lines were present in the spectra of December 12.
One year after the outburst, the nova was found to be well
into the coronal phase with the detection of [Si VI], [Si VII],
[Ca VIII], [S VIII], and [S IX] lines in its spectrum (Rudy et
al. 2005). V574 Pup was observed by the Spitzer Space Observatory one year after the outburst revealing strong coronal
lines (Rudy et al. 2006). The spectroscopic observations by
Lynch et al. (2007), three years after the outburst, showed
the persistence of the coronal phase. Rudy et al. (2006) have
remarked that the presence of strong coronal emission lines
suggests similarity with a He/N nova.

\begin{figure}[!htp]
\center
\includegraphics[bb=99 315 542 662, width=5in, clip]{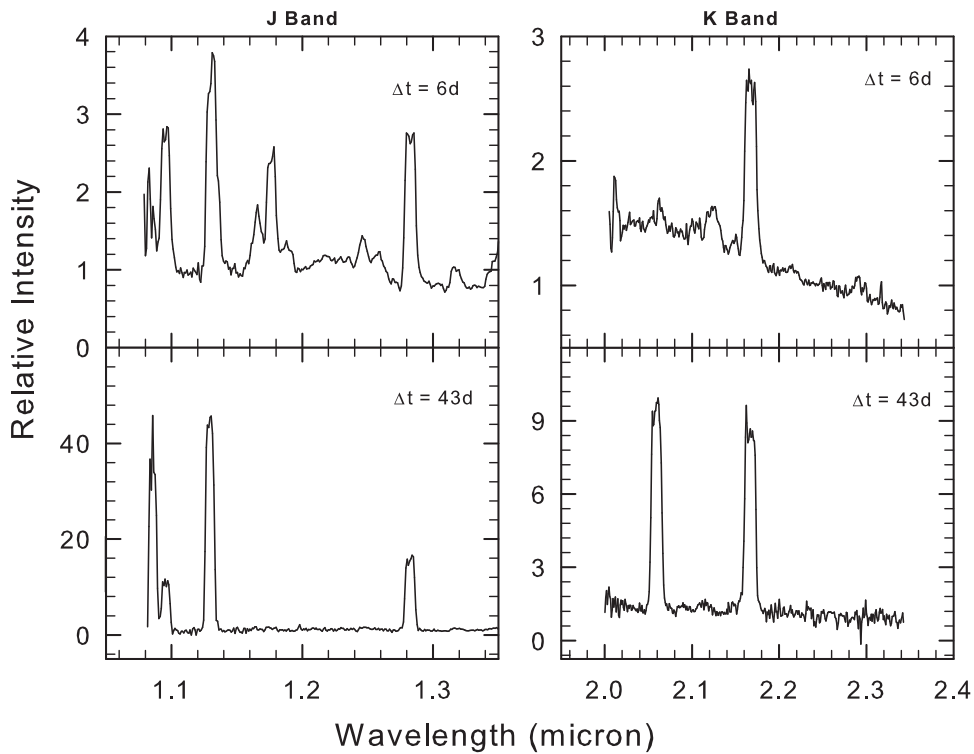}
\caption{The $J$ and $K$ band spectra of a Fe IIb class of nova (V574 Puppis). The change from Fe II to He/N class is illustrated  wherein the initial spectra show strong CI lines characteristic of the Fe II class (see Figures 1 and 3),  which quickly disappear to be replaced by lines more characteristic of He/N novae such as the very strong HeI lines at 1.0830 and 2.0581 $\mu$m lines. The time after outburst is indicated in the panels.}
\end{figure}

\section{Dust formation: near-IR signatures} Many novae form dust later during their evolution - the Fe II novae in particular show a stronger tendency to form dust. The formation of dust in the ejecta generally manifests itself by a sudden dip in the visible light curve  due to obstruction of the central source by the dust. This  is accompanied by a sharp increase in the IR emission as the dust particles absorb the visible radiation and are heated; the energy is subsequently reemitted in the IR.  The dip in the light curve  may not always happen - novae are known to produce optically thin dust shells too. As an illustration of this, Figure 6 shows the visible and near-IR light curves for three novae viz. V574 Pup, V1280 Sco and V476 Sct. In the case of V574 Pup, the nova did not form any dust. Hence the visible and near-IR light curves are seen to monotonically decline with time. V1280 Sco formed copious amounts of dust as evidenced by the sharp drop in its lightcurve about 12 days after maximum. However while the visible light curve showed this sharp dip, the $K$ band light curve maintained a steady brightness because of emission from the dust. In the N band (10 $\mu$m) the object was very bright at this stage ( Chesneau et al 2008). The source continues to evolve very slowly and even until recently was shrouded in a dusty bipolar nebula bright in the mid-IR (Chesneau et al. 2012). The bottom panel shows the light curve of V476 Scuti (unpublished data), a Fe II type nova, whose optical lightcurve did not show any dramatic dip due to dust formation. Yet the near-IR $H$ and $K$ band lightcurves showed a significant  increase in brightness, starting from  about 15 days after the outburst,  indicating the formation of dust. This was a case of the  dust being optically thin which cautions that dust formation may often go undetected if it is based solely on an anticipated deep dip
in the optical light curve as seen for e.g. in the case of nova V705 Cas (e.g. Evans et al. 1996)

The bottom right panel shows blackbody fits to the spectral energy distribution of V476 Sct derived using the $JHK$ magnitudes. The fits are made for different epochs and show the dust to be cooling rapidly with time. In the absence of longer wavelength data, we have sometimes used such blackbody fits to the SED of dust-forming novae to compute the mass and temperature of the dust using certain assumptions for the grain composition and grain size as described for example in  Woodward et al. (1993). However, better estimates of the dust parameters and properties  can be made by modeling the observed SED using more sophisticated radiative transfer codes like DUSTY (e.g. Evans et al. 2005, Evans et al. 2012).

The physical conditions necessary for dust formation have been addressed  on different occasions  but it is still not completely understood why some novae form dust and others do not (Gehrz 1988, Evans $\&$ Rawlings 2008). We addressed this issue while studying the case of the dust forming nova V1280 Sco and used LTE analysis to generate synthetic spectra to compare with those observed.
It was also found that use of a single temperature for the emitting gas  fails to
simultaneously reproduce the observed strengths of lines from all the elements.
Though a complex interplay between the Saha and Boltzmann equations is involved,
the principal reasons for this failure  can be intuitively traced to the
considerable diversity
in ionization potentials (IP) of the elements contributing to the spectrum. The
observed lines
in the spectrum  are from neutral species. If a higher temperature
of the ejecta is considered ($\sim$ 10,000K), then the ionization equation indicates that elements
like Na and Mg with low IPs of 5.139 and 7.646 eV respectively have a larger fraction
of their atoms in higher stages of ionization and very few in the neutral state. Therefore the
lines from the neutral species of these elements are  found
to be extremely weak - the reduction in strength is
compounded by the additional fact that they have low abundances. In comparison
to Na and Mg,  higher temperatures favor
a relatively larger fraction of neutral species for H, C, N and O  because of
their  significantly higher IPs (13.6, 11.26, 14.53 and 13.62 eV
respectively). On the other hand, instead of a high temperature, the use of only a
single lower temperature ($\sim$ 3500-4500K) enhances
the strength  of Na and Mg lines to an unacceptable extent - they  become too strong
vis-a-vis lines of other elements. Thus it becomes necessary to consider the possibility of
temperature variation in the ejecta. We  adopted the simplest
 scenario  viz. there exists  a hot zone in the ejecta outside which lies a relatively
 cooler  region. We conclude that the presence  of  lines of particularly
Na  and also Mg,  associated as they are with low excitation and ionization
conditions,   necessarily implies the  existence  of a cool zone.  Such regions could be  associated with clumpiness in the nova ejecta which likely provide the cool dense sites needed for molecule and dust formation. It was also corroborated statistically that whenever these lines are seen,
dust did indeed form in the nova (the novae we could identify which showed these lines  were V2274 Cyg (Rudy et al. 2003), V1419 Aql (Lynch. et al. 1995),  NQ Vul  (Ferland et al 1979), V705 Cas (Evans et al 1996),
 V842 Cen (Wichmann et al. 1991), V1280 Sco (Das et al. 2008), V476 Sct (unpublished data), V5579 Sgr (Raj et al. 2011) and V496 Sct (Raj et al. 2012)).

\input{epsf}
\begin{figure}[hbtp]
  \vspace{9pt}

  \centerline{\hbox{ \hspace{0.0in}
    \epsfxsize=2.6in
    \epsffile[2 1 341 333]{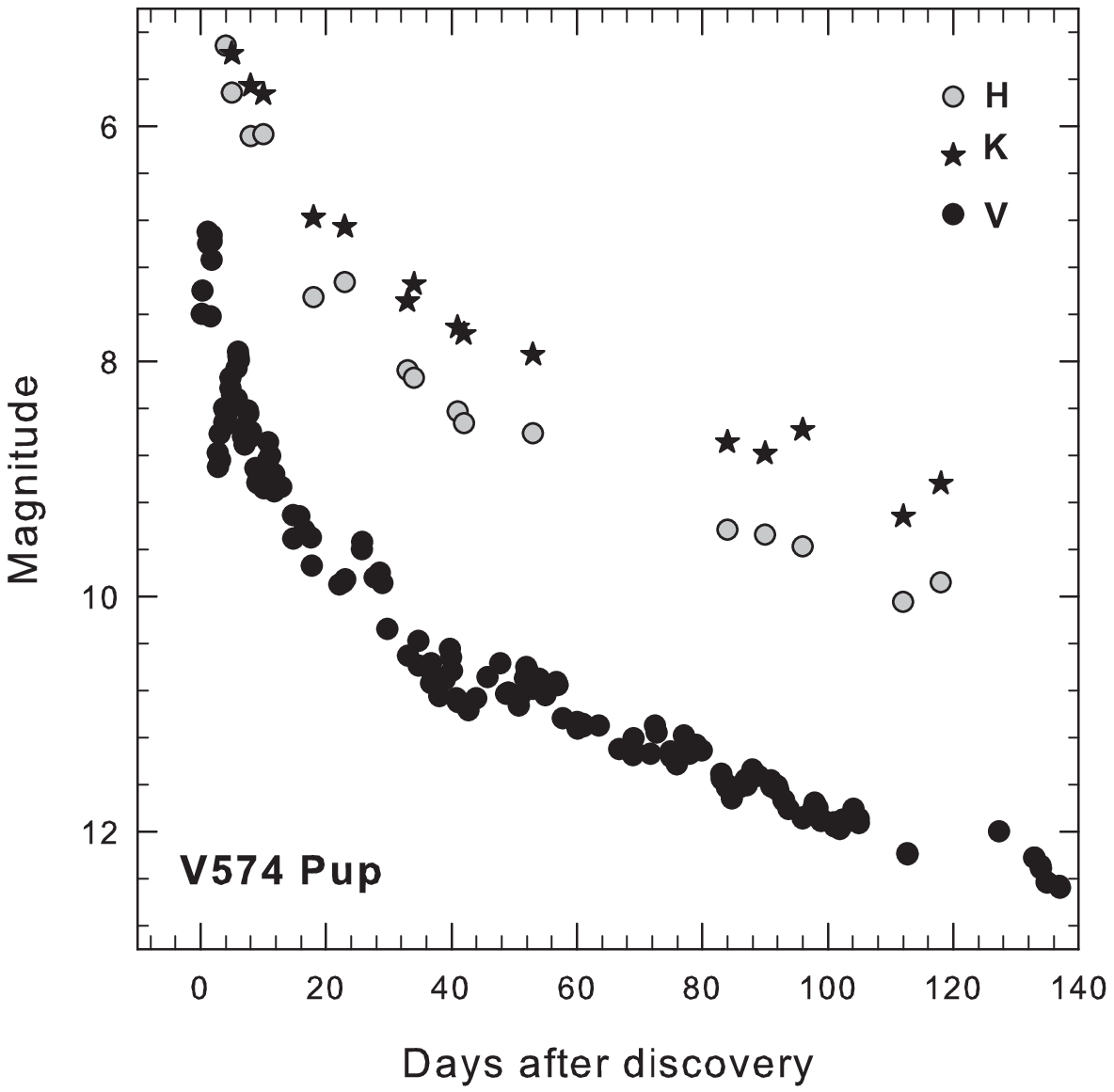}
    \hspace{0.0in}
    \epsfxsize=2.6in
    \epsffile[2 1 302 306]{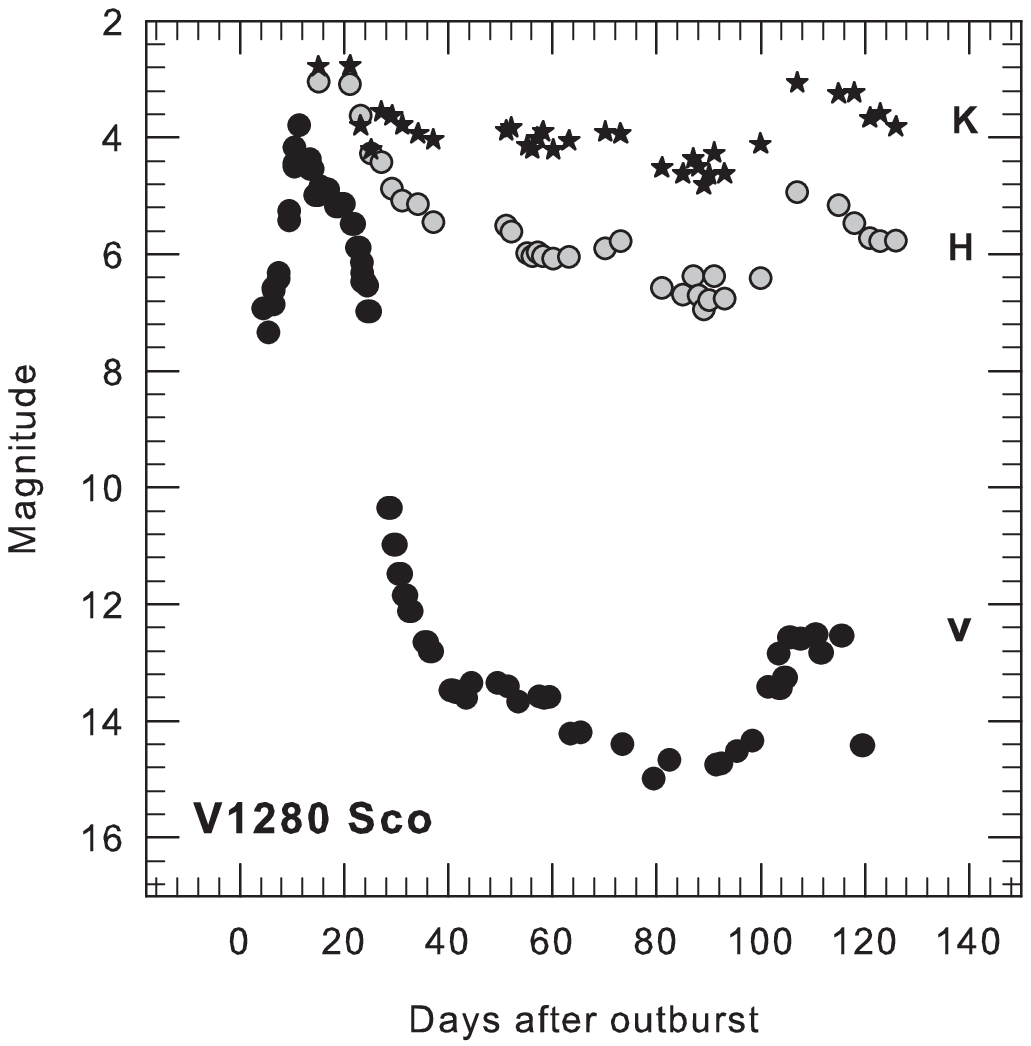}
    }
  }

  \vspace{6pt}
  \vspace{9pt}

  \centerline{\hbox{ \hspace{0.0in}
 \epsfysize=2.5in
 \epsffile[3 2 336 336]{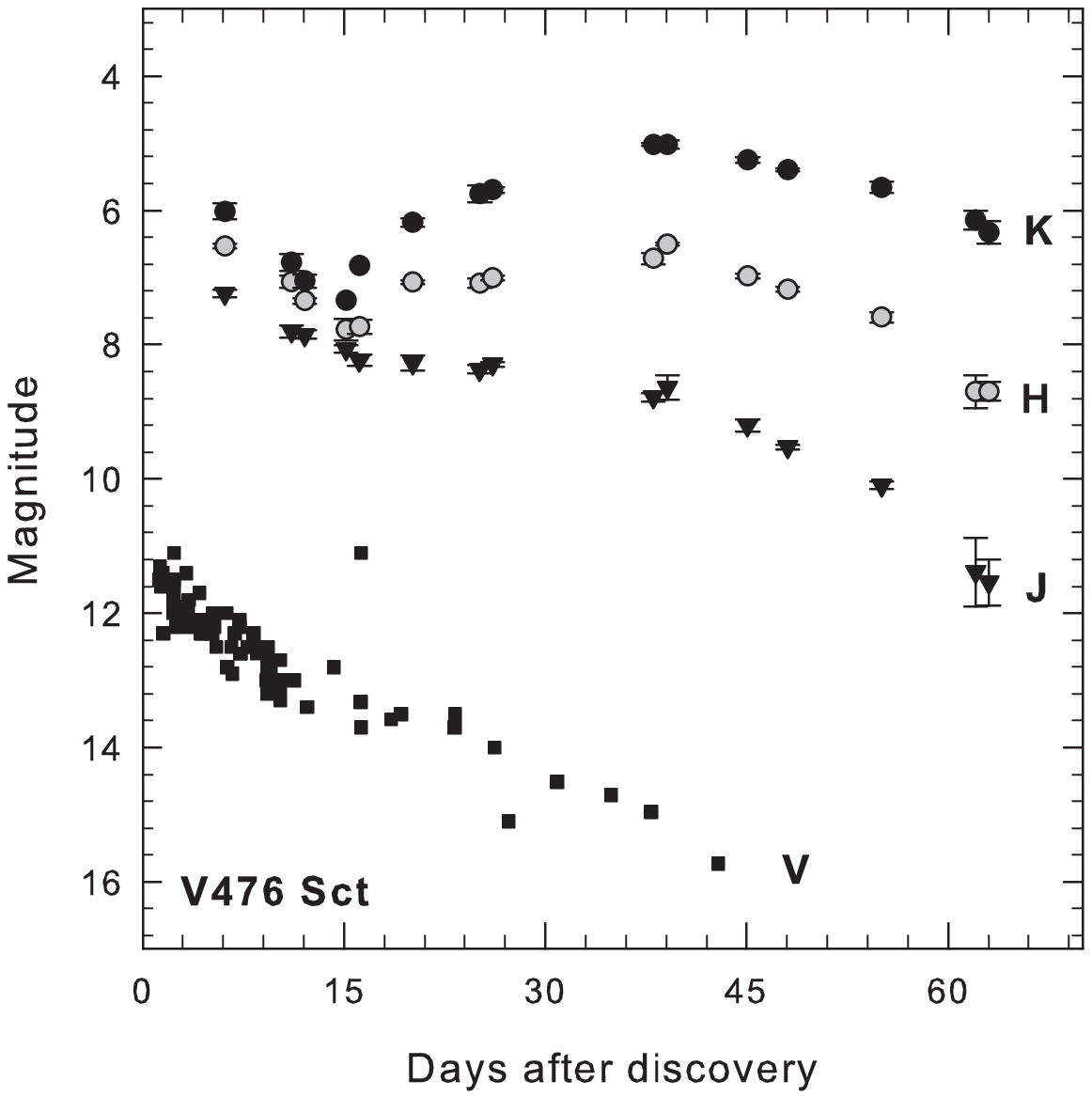}
  \hspace{0.25in}
    \epsfxsize=2.6in
    \epsffile[1 1 429 417]{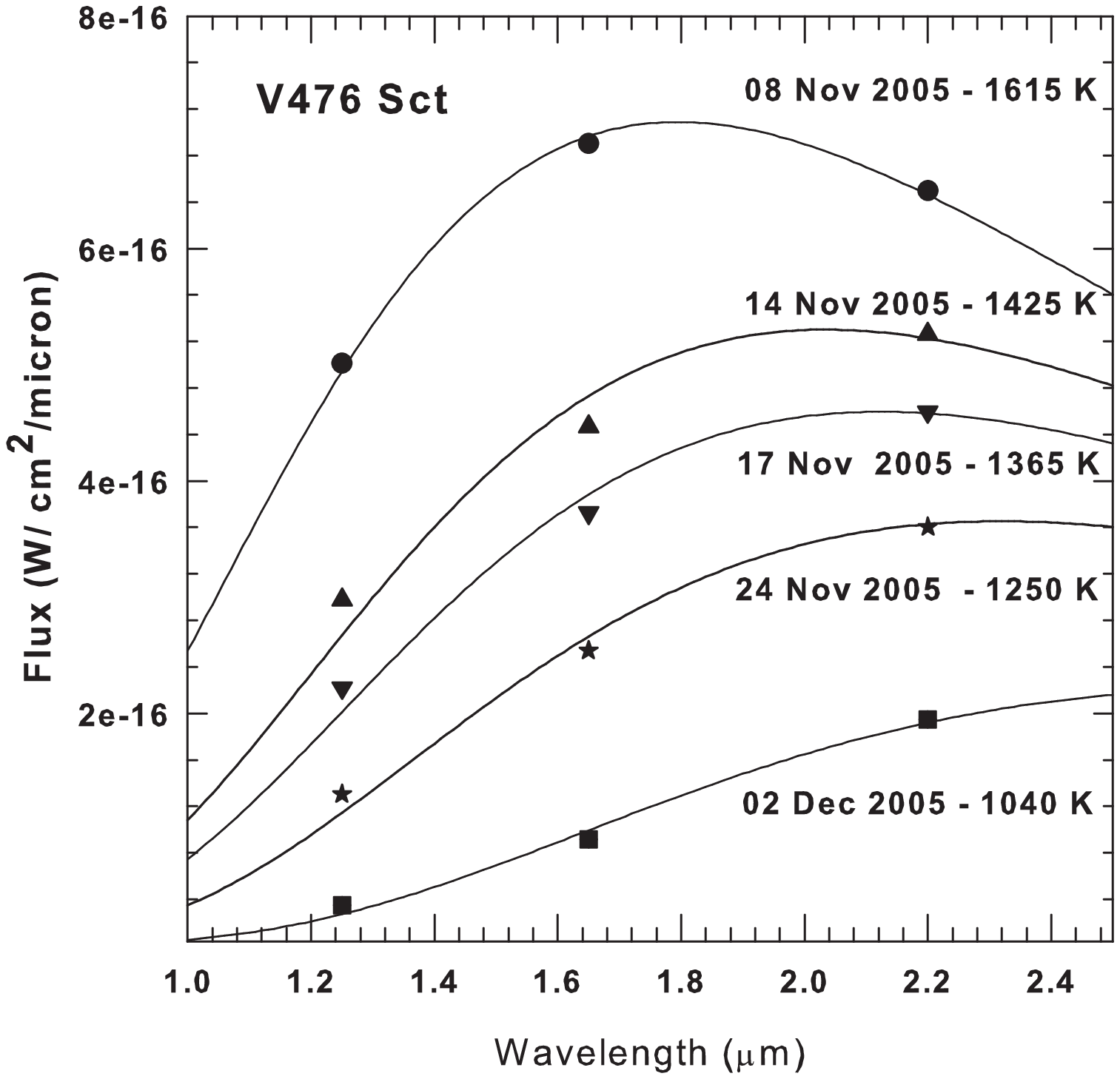}
    }
  }

  \vspace{6pt}
  \vspace{8pt}

  \caption{ A variety of optical plus IR light curves (LCs) illustrating the effects of dust formation on the LCs (a) V574 Pup:  a nova with no dust formation  showing the $V$ (dark circles) and $H$ $\&$ $K$ magnitudes (grey circles and star symbols) simultaneously decreasing with time (b) V1280 Sco, a nova with copious dust formation that creates a dip in the optical LC whereas the $H$ $\&$ $K$ band remains bright because of the dust emission (c) V476 Scuti - a nova with optically thin dust causing no dip in the optical LC (more discussion is the text) (d) Blackbody fits to the dereddened $JHK$ fluxes of V476 Scuti at 5 epochs showing the cooling of the dust as the peak of the SED shifts towards longer wavelengths with time.}
  \label{sub-fig-test}
\end{figure}

 \section{CO emission in novae} The fundamental band of molecular CO lies in the thermal $M$ band at 4.67 $\mu$m and hence is not accessible to us for observations. Observing in the $M$ band is anyway challenging because of the strong sky background. On the other hand the first overtone CO  bands are convenient to study being situated in the $K$ band at 2.29 $\mu$m and beyond. In novae, first overtone CO detections are very few - these are in V2274 Cyg (Rudy et al. 2003), NQ Vul (Ferland et al. 1979), V842 Cen ( Hyland $\&$ Mcgregor 1989, Wichmann et al. 1990, 1991), V705 Cas (Evans et al. 1996); V1419
Aql (Lynch et al. 1995), V496 Sct (Raj et al. 2012, Rudy et al. 2012a) and Nova Ophiuchi 2012 (Rudy et al. 2012b).
One of the biggest attractions in  a CO detection is the possibility of determining with accuracy the $^{13}$C yield in novae and thereby confirming the underlying nucleosynthesis predictions and theory. Novae have very distinctive yields of elements and for some  specific isotopes (like $^{13}$C, $^{15}$N and $^{17}$O) novae are believed to synthesize and contribute a significant fraction of the total galactic budget of these isotopes (Jose $\&$ Hernanz 1998, Gehrz et al. 1998). In the case of $^{13}$C specifically, the yield is expected to be high; the predicted values of the $^{12}$C/$^{13}$C ratio is $\sim$ 1 or more (more details on the predicted yields follow shortly).  This large expected yield makes it observationally simple, in principle, to accurately determine the $^{13}$C content  through measuring the strengths of  $^{13}$CO and $^{12}$CO bands. Unfortunately there are complications and the exercise generally allows only a limit to be set on the $^{12}$C/$^{13}$C ratio.

\subsection{Evolution of the CO emission: model predictions vs. observations} The detection of CO is also important in studying the astrochemistry of molecules in nova outflows and its role in dust formation. The earliest theoretical studies of the chemistry of novae ejecta were done by Rawlings (1988)   in the form of pseudo-equilibrium chemical models of the pre-dust-formation epoch. These models, which were developed with the main aim of explaining the observed presence of CO in novae, found that the outer parts  of the ejecta have to be substantially more dense and less ionized than the bulk of the wind for substantial molecule formation to occur. For this to occur carbon has to be neutral. In a neutral carbon region, the carbon ionization continuum, which extends to less
than  1102 $\AA$, shields several molecular species against the dissociative UV flux from the central star. An extended and  more refined  model for molecule formation in the nova outflow in the early stages  is presented in Pontefract $\&$ Rawlings (2004; hereafter PR; also see Evans $\&$ Rawlings 2008).

A significant result in PR is the prediction of the evolution of the
fractional CO abundance with time. Two models are considered -  model A considers oxygen rich ejecta and model B considers carbon rich ejecta. Figures 1 and 2 of PR show the evolution of the fractional abundance of different molecules and radicals, including CO,  with time. It is seen that in both models,  the CO abundance remains constant upto about 2 weeks after
outburst ($\sim$ 12 days in case A and $\sim$ 15 days in case B). This behavior,
and the length of its duration, seems to be generic to the models. During this phase the CO is saturated - that is to say, all the available oxygen or carbon, whichever has the lower abundance, is completely incorporated into forming
CO. After this there is a sharp decline in the CO abundance as CO is destroyed mostly by
reactions with N and $\rm N^+$. During this stage, from Figures 1 and 2 of PR, we estimate an approximate decrease in CO  by a factor of 1000 in $\sim$ 27days for model A and a decrease by a factor of 100 in $\sim$ 16 days for model B.

While the earlier detections of CO were limited to one or two epochs of observation, CO in V2615 Oph (Das et al 2009) and also V496 Sct (Raj et al 2012) was detected over a significant duration of time presenting an opportunity to study  the  formation and destruction of CO during a  nova outburst. We restrict ourselves to V2615 Oph here and show in Figure 7 the mosaic of $K$ band spectra showing the evolution of the CO.
Figure 8 shows  model fits to the observed CO bands on selected epochs to estimate the CO mass, temperature and $^{12}$C/$^{13}$C ratio.
The present observations and modeling show that the CO mass was  constant between 28 March and 3 April i.e. for a period of 7 days. During this phase,
the gas temperature and mass are found to be fairly constant  in the  range of  4000 - 4300K and
2.75x$10^{-8}$  to  3.25x$10^{-8}$ $M_{\odot}$ respectively. This puts a lower limit on the duration of the saturated phase  since our observations  commenced on 28 March, nearly 8-9 days after the beginning of the outburst on March 19.812 UT. It is possible that the CO emission was present, and at  similar levels, between March 19 and March 28 also. After all, CO has been seen very early after commencement of the eruption as in the case of V705 Cas (Evans et al. 1996). If that is the case, then the upper limit on the saturated-phase timescale would be around 15 days. Thus the evidence indicates that a phase does exist when the CO mass is constant and whose duration $t_s$ is constrained within 7 $\le$ $t_s$ $\le$ 16 days. This observational finding is consistent with the predicted timescales of $\sim$ 12-15 days from the PR model calculations. Between 3 April and 7 April there is a sudden decrease in the CO strength. This quantitative behavior i.e. the rapid destruction of CO following the saturated phase  is again
largely consistent with the predictions  of the theoretical model (PR). More discussion on this is available in Das et al (2009).

However one needs to be cautious in drawing any strong  conclusion based solely on one nova. One should also consider carefully the evolutionary behavior of the CO in the few other instances where more than one epoch of observations/detections of CO have been recorded such as in V496 Sct (Raj et al. 2012, Rudy et al. 2012a) and V705 Cas (Evans et al. 1996).

\begin{figure}[!htp]
\center
\includegraphics[bb=2 0 694 760, width=5in,height=5in, clip]{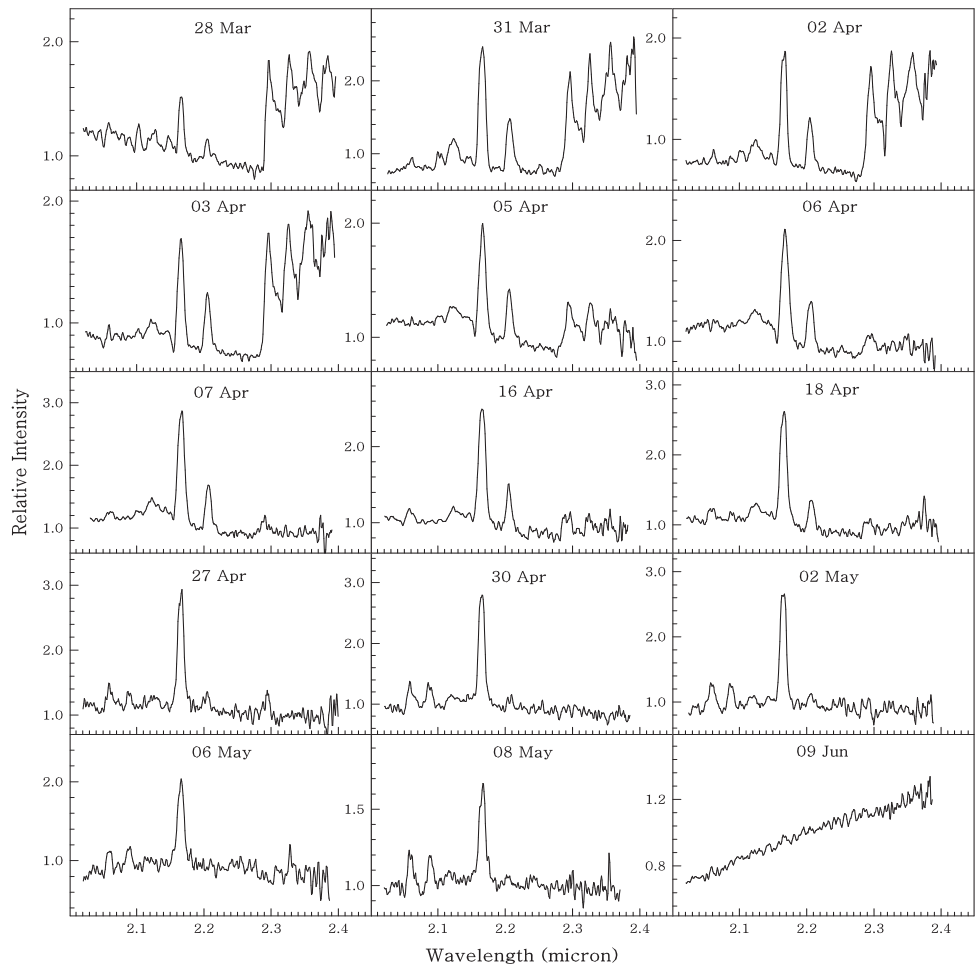}
\caption{The $K$ band spectra of V2615 Oph on different days in 2007 (discovery date 2007 March 19.812 UT) with the flux normalized to unity at 2.2 ${\rm{\mu}}$m. The evolution of the strength of the CO bands with time discussed in see section 5.1 can be seen. The reversal of the slope of the continuum on the last  date is due to an IR excess from dust formation.}
\end{figure}

\begin{figure}[!htp]
\center
\includegraphics[bb= 0 0 304 620, width=4in,height=5.75in, clip]{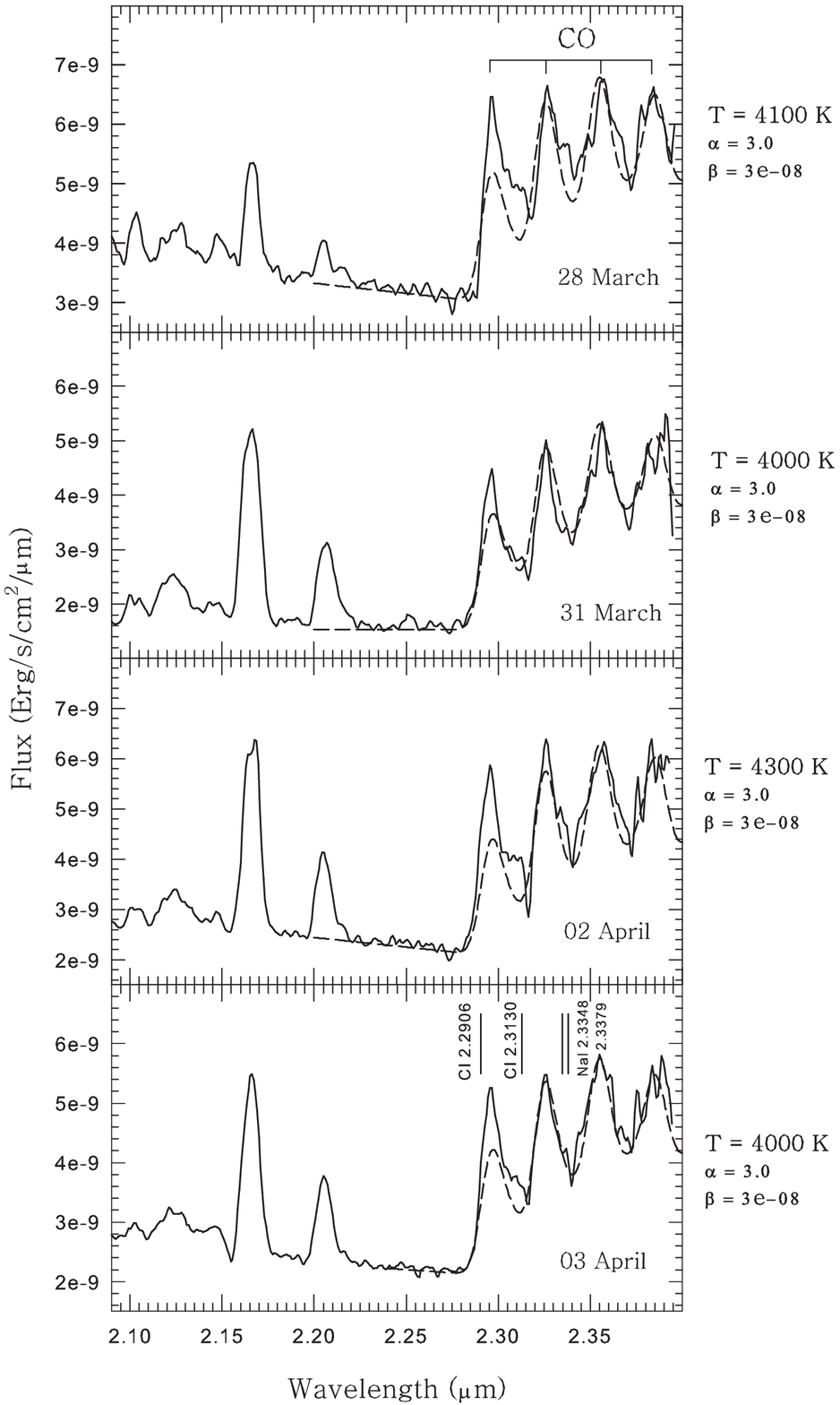}

\caption{Model fits (dashed lines) to the observed first overtone CO bands
in V2615 Oph. The fits  indicate  a constant  CO mass
of  3.0x$10^{-8}$ $M_{\odot}$ on all  days while the temperature of the gas
$T_{CO}$ is estimated to be  4100, 4000, 4100 and 4300K (with an error
of $\pm$ 400K) for 28 $\&$ 31 March and  2 $\&$ 3 April 2007 respectively. The bottom panel
shows the marked positions of certain CI (at 2.2906 and
2.2310 $\mu$m) and NaI (at 2.3348 and 2.3379 $\mu$m)lines that complicate the modeling.
The other prominent lines seen in the spectra are Br $\gamma$ at
2.1655 $\mu$m; NaI 2.2056, 2.2084 $\mu$m blended with weaker emission from CI 2.2156, 2.2167 $\mu$m
lines and other CI lines between 2.1 and 2.14 $\mu$m. The position of the $^{12}$C$^{16}$O bandheads are shown in the top panel}
\end{figure}

\subsection {Dust formation and CO emission}
Does CO formation always precede or accompany dust formation in a nova?  It is well established that all novae which showed CO emission proceeded  to form dust. However,  the converse may not be true i.e. all dust producing novae may not produce CO; atleast not significant amounts of it. This is based on our observational experience wherein we have witnessed dust production in novae but seen no evidence of CO emission. V1280 Sco is a good example of this. It is possible that  CO emission was present in V1280 Sco but at too low a level for us to detect it. But that appears unlikely because V1280 Sco was a bright source and we obtained good S/N spectra of it. Our extensive and regular monitoring of the object, starting from $\sim$  3 days past visual maximum,  also makes it unlikely that CO production was present in the nova but  was short-lived and hence eluded detection.

\subsection{ The $^{12}$C/$^{13}$C ratio in classical novae}

The $^{12}$C/$^{13}$C ratio can be estimated from model fits to the observed CO bands as shown in Figure 8 and described in details for e.g. in Das et al (2009) for the case of V2615 Oph. While making the fits we note that the CO bands are almost certainly  contaminated by CI lines at 2.2906 and
2.2310 $\mu$m and possibly from NaI lines at 2.3348 and 2.3379 $\mu$m. This introduces a considerable complication. The position of these lines are marked in the bottom panel of Figure 8 and there are discernible structures at these positions,  in most of the profiles, indicating these lines are present. The 2.29 $\mu$m ($v$ = 2-0) band  would appear to be  affected, especially by the CI 2.2906 line which can be significant in strength in Fe II type novae and whose presence is responsible for lack of agreement between model and observed data in this region. In view of the above, one has to rely more on the fits to the higher bands ( the $v$ = 3-1 and 4-2 bands) while estimating the CO parameters.  One has also to assume  the CO emission to be optically thin. A major  complexity is that each rotational line within a band will be broadened considerably because of the high velocity dispersion in  the nova ejecta (typically 1000km/s or more). Thus the $^{13}$CO and $^{12}$CO bands will tend to overlap because of Doppler broadening and will not be seen  clearly as distinct components. This is in contrast to what is routinely seen in the spectra of cool stars where the $^{13}$CO and $^{12}$CO components can be easily discerned by the eye even when the ratio of $^{12}$C/$^{13}$C in such stars is much higher than that predicted in novae.  Because of these complexities, one is unfortunately able to set only a limit on the $^{12}$C/$^{13}$C ratio in novae. The available observed values of this ratio are presented below and compared with predictions.

For  NQ Vul,  Ferland et al. (1979) from observations 19 days after  outburst or 20 days before the
large visual fading associated with dust formation estimated a limit of $^{12}$C/$^{13}$C greater than 3.0. In V2274 Cyg, Rudy et al. (2003)  from observations 17 days after discovery, determined  $^{13}$C/$^{12}$C $\ge$
0.83 $\pm$ 0.3.  In nova V705 Cas, observations by Evans et al. (1996) taken at two epochs i.e. one day before maximum light and 26.5 days after maximum light yielded estimates for the $^{12}$C/$^{13}$C ratio to be $\ge$ 5. In DQ Her this ratio was found to be $\ge$ 1.5  by Sneden and Lambert (1975) while in V842 Cen observations between days 29 and 45 after outburst by Wichmann et al. (1991) showed the $^{12}$C/$^{13}$C ratio to be 2.9 $\pm$ 0.4. Our observations show the $^{12}$C/$^{13}$C to be $\ge$ 2 in V2615 Oph (Das et al. 2009)and $\ge$ 1.5 in V496 Sct  (Raj et al. 2012) though for V496 Sct Rudy et al (2012a) estimate the value to be 1.25.

    The expected $^{12}$C/$^{13}$C ratio in the ejecta of a nova  has been computed (e.g. Jose $\&$ Hernanz (1998); Starrfield et al. (1997)) and shown to be dependent on the white dwarf mass among other parameters. Considering the novae with the carbon-oxygen core white dwarfs,  different models by Jose $\&$ Hernanz (1998) show that this ratio is approximately constrained between 0.3 to 0.65 for a white dwarf mass between 0.8 to 1.15 $M_{\odot}$; Starrfield et al. (1997) find the $^{12}$C/$^{13}$C ratio to decrease from 2.4 to 0.84 as the white dwarf mass increases form 0.6 to 1.25 $M_{\odot}$. There may be  no major conflict between observed and predicted values of this ratio but the majority of the observed values of the $^{12}$C/$^{13}$C ratio are greater than unity (whereas a value less than 1 is generally expected)  leaving a small doubt whether  $^{13}$C is really being  synthesized to the high levels predicted by the theoretical models. However, as mentioned earlier, there is scope for error in estimating this ratio from observations.

\section*{Acknowledgments}

The research work at the Physical Research Laboratory, Ahmedabad  is funded by the Department
of Space, Government of India.

\end{document}